# Multiple ferroelectric nematic phases of a highly polar liquid crystal compound

Rony Saha[1,2], Pawan Nepal[3], Chenrun Feng[2,4], Md Sakhawat Hossein[4], James T. Gleeson[1], Samuel Sprunt[1,2], Robert J. Twieg*[3], Antal Jákli*[1,2,4]

[1]Department of Physics, Kent State University, Kent, OH 44242, USA

[2]Advanced Materials and Liquid Crystal Institute, Kent State University, Kent, OH 44242, USA

[3]Department of Chemistry and Biochemistry, Kent State University, Kent, OH 44242, USA

[4]Materials Science Graduate Program, Kent State University, Kent, OH 44242, USA

*: corresponding authors: rtwieg@kent.edu, ajakli@kent.edu

**Abstract:**

*Ferroelectric nematic liquid crystals represent not only interesting fundamental science, but they also hold promise for storage capacitors with high power density or new information display technology having sub-millisecond switching. In this work we describe the synthesis and measurements of the physical properties of a new highly polar ferroelectric nematic compound, 4-nitrophenyl 4-[(2,4-dimethoxylbenzoyl)oxy]-2-fluorobenzoate (RT11001). The dipole moment of this material (along the long molecular axis) is calculated to exceed 11.5 Debye. We employ a wide range of physical characterization methods including differential scanning calorimetry (DSC), mass density measurement, optical birefringence, polarizing optical microscopy (POM), electric current analysis, and electro-optical switching, to show that RT11001 has three distinct ferroelectric states, $F_1$, $F_2$ and $F_3$. $F_1$ is purely orientationally ordered ferroelectric nematic phase ($N_F$), $F_2$ has a ferroelectric nematic with possibly short-range hexagonal order normal to the director ($N_{hF}$), and we conjecture that $F_3$ has a long-range hexagonal order normal to the director ($Col_{hF}$).*

1. Introduction

Uniaxial nematic liquid crystal materials are anisotropic fluids with a single optical axis; this axis may be reversibly reoriented under the influence of an AC electric field when the dielectric tensor is also anisotropic. The ability to change optical properties with a small electric field has made nematic liquid crystals the dominant technology for electronic information display, such as today's flat panel displays. Over the last 130 years, several other nematic liquid crystal phases, such as chiral[12], biaxial[3] and twist-bend nematic phases[5] have been both predicted and observed. Although in bent-core nematic[4] materials there are indications of the presence of microscopic ferroelectric smectic clusters[6], until recently no nematic liquid crystals have been found to be macroscopically polar. This is surprising given that the first mean-field theory of the nematic-isotropic transition (proposed by Born in 1916) envisioned strong dipole-dipole interactions leading to polar ordering in nematic liquid crystals.[7] It was later predicted that moderate electric dipole interactions between disk-shaped molecules may be sufficient to induce a ferroelectric nematic phase.[8] However, there were no unambiguous experimental indications of a fluid ferroelectric nematic phase until Nishikawa *et al*[9] reported a polar nematic phase formed by a rod-shaped molecule having a large dipole moment; this phase showed a polarization up to 5 μC/cm² and dielectric constant of $\varepsilon \sim 10^4$ at 1 kHz. In addition , Mandle *et al* reported a nematic-nematic transition in a thermotropic compound (referred to as RM734) having molecular dipole moment of about 10 Debye.[10,11] The lower temperature nematic phase was subsequently identified as the splay nematic phase[12], with the splay deformation being the result of polar molecular ordering and with a dielectric constant approaching $\varepsilon \sim 10^4$ at 100Hz.[13] More recently, Chen et al, proposed that this compound's polar nature does not induce a spontaneous splay deformation, but rather that the lower temperature phase is a uniform ferroelectric nematic phase with a spontaneous polarization as high as 6 $\frac{\mu C}{cm^2}$.[14]

Ferroelectric nematic liquid crystals are not only extremely interesting from the point of view of basic science, but they are also promising for high energy and high power density capacitors[15] and for novel information displays capable of sub-millisecond electro-optic responses.[16] All of these seminal developments make the synthesis and characterization of a wide-range of new ferroelectric nematic liquid crystal materials very significant.

A widely studied class of compounds exhibiting multiple nematic phases (including the ferroelectric nematic phase) are 4-nitrophenyl 4-[(2,4-dialkoxybenzoyl)oxy]benzoates, which contain three benzene rings joined by two ester groups.[10] The phenyl terminal ring in these molecules is often substituted with a strong electron withdrawing group such as nitro, while the other terminal ring is often substituted with donor groups such as methoxy. In this class, numerous variations of substituents on the two terminal rings have been described, but much less attention has been paid to modification of the central benzene ring.[15] As such, we decided to concentrate our initial design and synthesis effort related to nematic ferroelectric substances on modifications in this center benzene ring.

In this work we describe the results of the introduction of a single fluorine adjacent to the ester carbonyl group of the center benzene ring of RM734 to give 4-nitrophenyl 4-[(2,4-dimethoxylbenzoyl)oxy]-2-fluorobenzoate (RT11001). In addition to the description of the synthesis of RT11001, we also describe the characterization of critical physical properties. The lateral fluorination at this site in the center ring produces some of the anticipated effects on phase transitions and simple physical properties, such as the dipole moment, relative to RM734. In addition, we find as a result of this simple structure change that RT11001 has at least two ferroelectric nematic phases and, below those, a third ferroelectric phase that may also possess positional order perpendicular to the orientational order.

## 2. Material synthesis

The thermotropic compound RT11001 was synthesized in four steps as outlined in Scheme 1 and generally follows the route prescribed by Mandle[10] with minor modifications.

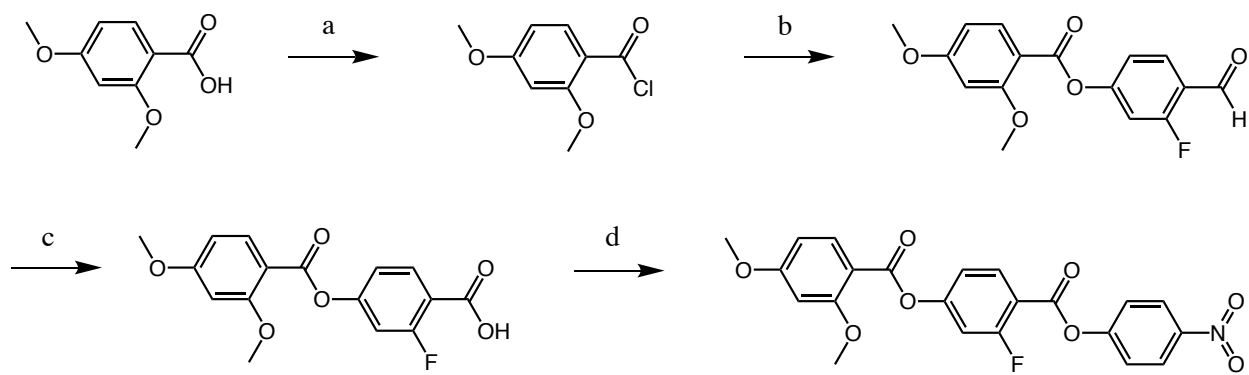

a) (COCl)$_2$, CH$_2$Cl$_2$ (quantitative)    b) 2-fluoro-4-hydroxybenzaldehyde, pyridine, CH$_2$Cl$_2$ (88%)

c) Oxone, DMF (quantitative)    d) 4-nitrophenol, DCC, DMAP, CH$_2$Cl$_2$ (36%)

*Scheme 1. Synthesis route with reagents and yields for RT11001.*

Commercial 2,4-dimethoxybenzoic acid was converted to the acid chloride with excess oxalyl chloride. This acid chloride was reacted with commercial 2-fluoro-4-hydroxybenzaldehyde to provide the two-ring phenyl benzoate ester bearing the aldehyde group. In the subsequent step we chose to oxidize the aldehyde to the carboxylic acid with Oxone.[17] All of these initial steps were efficient and did not require any significant intervention with crystallization or chromatography. In the final step the two-ring phenyl benzoate ester now bearing a carboxylic acid group was esterified with 4-nitrophenol. Amongst the many options available[18,19] to perform the esterification we chose general Steglich conditions (DCC/DMAP) to create final the three-ring diester. At this stage chromatography and recrystallization were required in order to obtain the final product in a high state of purity. The spectroscopic data obtained for the intermediates and the final product (see ESI) all concurred with expectations and comparisons with similar reported structures.

3. **Results and discussion**

A. *Thermal properties*

Figures S1 and S2 of Electronic Supplementary Information (ESI) show the DSC curves of RT11001 during heating and cooling at a rate of 5º C/min. In Figure S2, the sample was recrystallized from 1-propanol and kept in a vacuum oven (100 Torr, at room temperature for three days). At room temperature, RT11001 is crystalline; during heating, two distinct peaks were observed: an exothermic peak at 91.73 ºC ($\Delta H$ = 5.33 J/g), and an endothermic peak at 161.88 ºC ($\Delta H$ = 93.2 J/g). During cooling, six distinct peaks were observed. The first transition observed at 162.96 ºC ($\Delta H$ = 1.31 J/g) corresponds to the transition to a nematic phase; the other peaks observed on cooling are at 140.54 ºC ($\Delta H$ = 2.41 J/g), 114.63 ºC $\Delta H$ = 1.29 J/g, 108.65 ºC $\Delta H$ = 1.18 J/g, 93.33 ºC ($\Delta H$ = 36.04 J/g) and 69.84 ºC ($\Delta H$ = 8.52 J/g). DSC scans of RT11001 samples that were not kept under vacuum prior measurements reproduced most of the observed peaks with slight changes of the temperatures and enthalpy values, except that the two peaks at 114.63 °C and 108.65 °C merged to one larger peak starting at around 112 ºC ($\Delta H$ = 5.722 J/g).

Representative polarized optical microscopy (POM) textures of RT11001 in a 10 μm thick cell with planar alignment are presented in Figure S3 of ESI for each of the phases identified by DSC. Upper images show the textures in heating. The left image represents a crystal phase at 58°C. In agreement with the exothermic peak in DSC, at 91.5°C there is a phase change to a low birefringence state shown in the middle image at 98°C. This texture does not change until 161°C, then it transitions to a uniformly aligned uniaxial nematic phase, as shown at 162°C in the top-right corner of Figure 1. Lower images show three representative textures below 140.5°C: at 128°C (right), 104°C (middle) and 94°C (left). All three are similar, showing stripes roughly parallel to the rubbing direction. The thickness of the walls separating the uniform birefringence stripes increases on cooling, being approximately 5µm, 10µm and 20µm at 128°C, 104°C and 94°C, respectively. Other than the width of the defect walls, the textures do not change upon decrease 70°C, at which point crystalline domains appear.

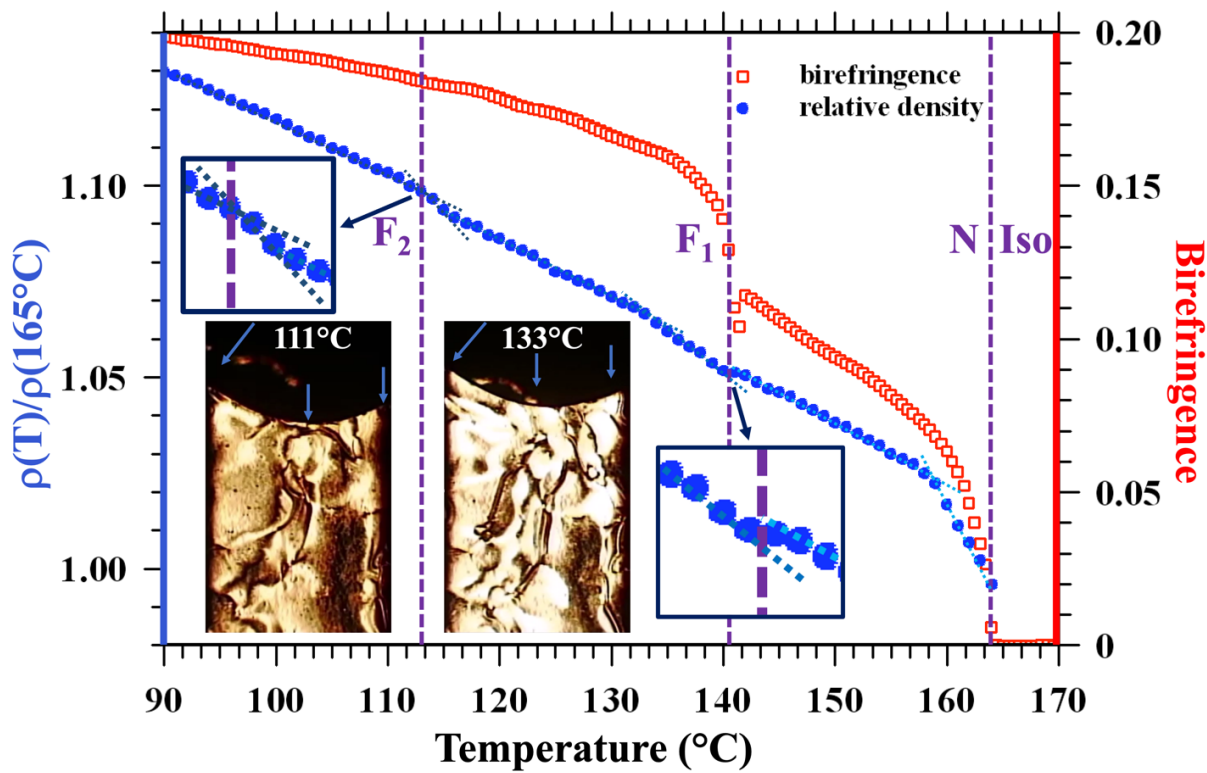

*Figure 1: Temperature dependence of the relative mass density ρ(T)/ρ(165°C) (blue dots) and of the birefringence (red squares). Inset shows POM textures near the meniscus in a 50µm rectangular capillary used for the density measurements at 133°C and 111°C.*

Figure 1 shows the temperature dependence of the birefringence and the relative mass density. As the temperature is decreased, $\Delta n$ increases from zero to 0.12 from 164°C to 140°C in the nematic range, then increases sharply up to 0.14 within a 1°C range below 140.5°C, then increases smoothly up to 0.2 at 90°C. No anomalous change is observed around 112°C, where DSC shows an additional phase transition.

The $\rho(T)/\rho(165°C)$ plot shows an abrupt increase below the I-N transition and then a fairly linear increase between 160°C and 140.5°C. At 140.5°C, where the DSC shows a phase transition, there is an abrupt increase in the slope of $\rho(T)$. The slope remains constant until approximately 132°C, then lessens slightly until about 112°C, where there is an abrupt increase, which corresponds to the second DSC peak starting at 112°C (see Figure S1) and which persists down to 110°C.

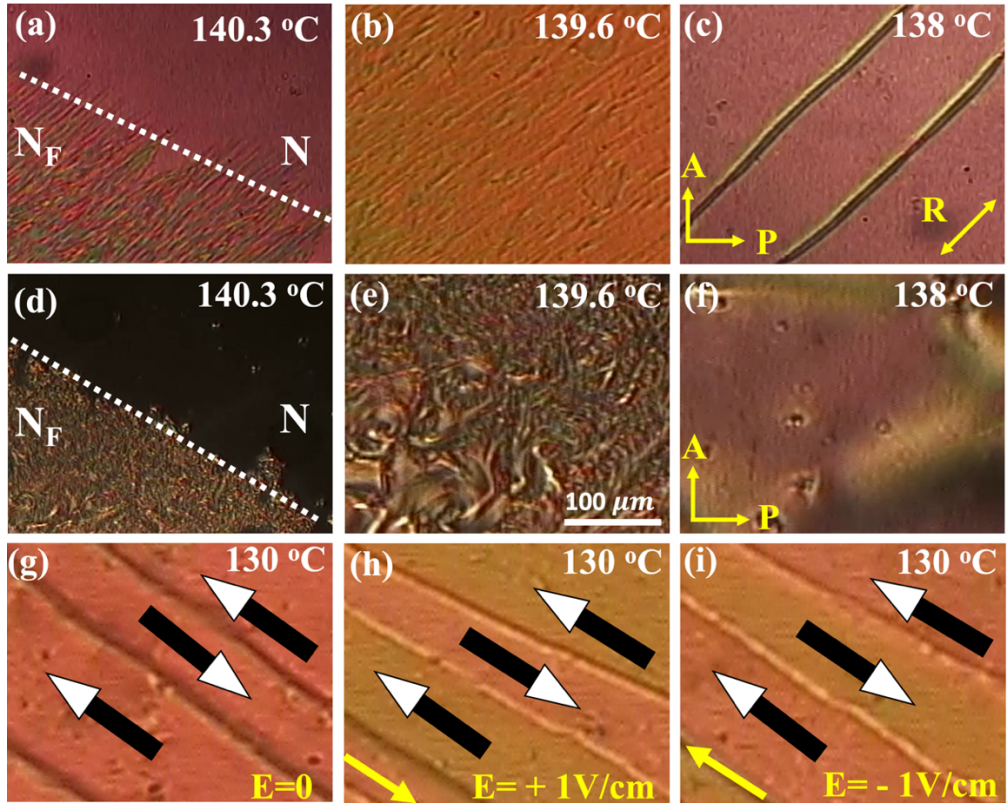

*Figure 2: POM textures of 10 μm cells of RT11001 at and a few degrees below the transition from the N phase after cooling from the isotropic state. (a-c) planar alignment (SE2170 without ITO) cooling rate: 0.5°C/min. Axis with double arrowheads indicate rubbing direction (R); (d-f) homeotropic alignment (SE5661 without ITO), cooling rate: 1°C /min. (g-i) POM textures of the film in (a-c) at 130°C under in-plane electric fields. Left: E=0. Uniform birefringence colors separated by about 10-15 μm thick defect walls, Middle and right: E=±1V/cm electric field applied in the direction of yellow arrows.*

Figure 2 presents POM textures of 10μm RT11001 films in the vicinity of the N-$F_1$ transition. Figure 2(a and d) represent textures at 140.3°C showing the first order N-$F_1$ transition with the uniform nematic texture at the top-right and the $F_1$ phase with grainy texture at the bottom-left. Note that in the planar cell the nematic phase has a uniform birefringent texture, while in a homeotropic cell the nematic texture is dark between crossed polarizers. The $F_1$ phase, however, has a grainy birefringent texture in both alignments, indicating that the homeotropic alignment is not effective in this phase. The textures in Figure 2(b and e) at 139.6°C show annealing of the grains to a more uniform texture with fine stripes parallel to the rubbing direction for the unidirectionally antiparallel rubbed planar cell (Figure 2(b)), while in the non-rubbed homeotropic cell the stripes have random directions (Figure 2(e)). The planar texture in Figure 2(c) at 138°C shows a fully annealed texture with about 100μm wide uniform domains separated by defect walls that are about 10μm thick. At 2°C below the N-$F_1$ phase transition, the optical path difference (estimated from the Michel-Levy chart) is $\Delta n \cdot d \sim 1600\ nm$, which corresponds to a birefringence of $\Delta n \sim 0.16$. This value is lower than that reported for RM734[14] (~0.22) but is in agreement with our direct birefringence measurements in Figure 1. The texture in the homeotropic cell in Figure 2(f) at 138°C shows a birefringence color comparable to that seen in Figure 2(c) for the planar cell, indicating degenerate planar alignment on the homeotropic substrates.

## B. Electric properties

The stripes separated by defect walls shown in Figure 2(c) are similar to those observed by Chen et al[14] where the stripes were described as domains with alternating directions of ferroelectric polarization. To verify this in RT1101, we applied a small ($\pm 10^{-4}$ V/μm) in-plane, DC electric field and compared the stripe textures. As seen in Figure 2(g), under zero field, adjacent stripes appear the same (similar to Figure 2(c)). When the field is applied, we see slight differences of the color between adjacent stripes; these changes alternate when the field is reversed. We note that this field is too small to induce a switching of the polarization direction. We attribute these observations to the enhancement/suppression of the polar order (and hence the birefringence) when the electric field is aligned/opposite with the polarization direction. In addition, under prolonged application of the DC field we also observe a slight reduction in the width of stripes having lower birefringence, indicating a slow realignment of the polarization; this realignment presumably takes place within the defect walls. We do not observe twisted domains such as those reported in planar-aligned samples of RM734.[16]

We measured the electric current responses of RT11001 in two geometries: First, using an in-plane electric field (as done for RM734 by Chen et al[14]), and second, with a potential difference applied normally between parallel, ITO coated substrates treated for planar alignment (which is the usual sandwich cell geometry employed in the characterization of ferroelectric smectic materials). [21]

In planar aligned cells, where the in-plane field was applied between 2 mm wide parallel ITO strips deposited on one plate and separated by 1 mm gap, even fields as small as $10^{-3}$V/μm could switch the polarization fully. Due to the inhomogeneous electric field around the electrodes, in this cell geometry we cannot deduce the dielectric constant and the electric conductivity from the capacitive and resistive currents. However, the net polarization charge $Q$ appearing on the in-plane electrodes is independent of the size of the electrodes and is determined only by the area $A = 1cm \times 10\mu m = 10^{-7}m^2$ (length of the electrodes multiplied by the film thickness) normal to the polarization vector, as $Q = P_o \cdot A$. For this reason, we used the in-plane cell geometry to measure the temperature dependence of the switching voltage, $V_s$, the saturated polarization, $P_o$ and the switching time, $\tau$.

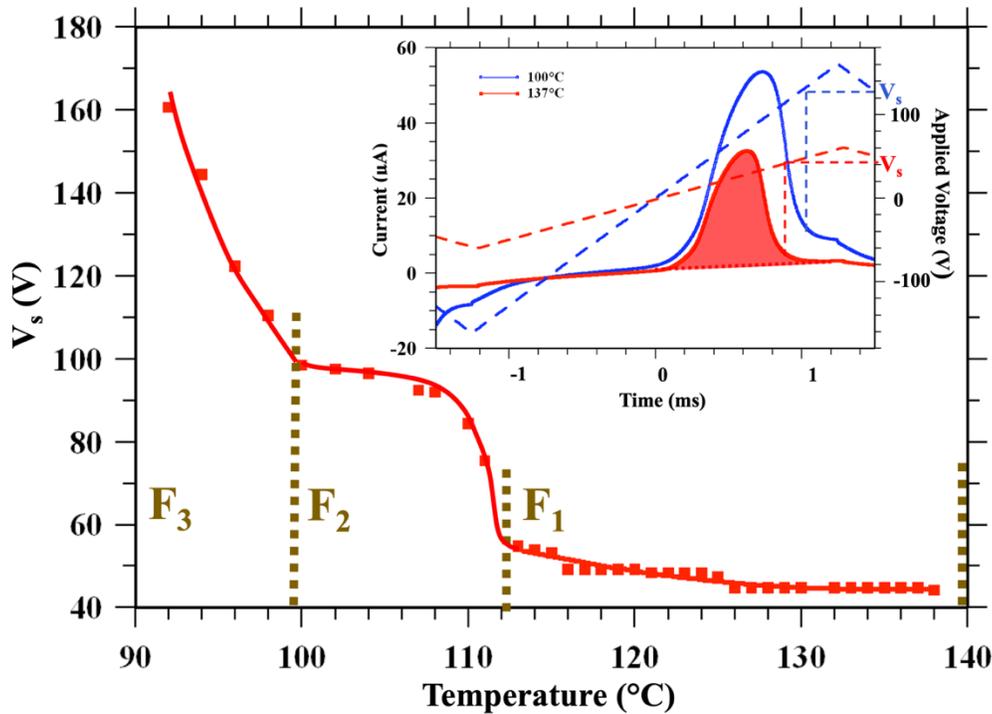

*Figure 3: The temperature dependence of the electric voltage $V_s$ where the switching of the polarization is completed (red squares). Inset: Time dependence of the electric current (left axis) and of the applied voltage (right axis) with the illustration of how $V_s$ and the polarization $P_o$ is determined from the shaded area.*

The electric current responses under various triangular waveforms applied in-plane are shown in the inset to Figure 3. The temperature dependence of the saturation voltage, $V_s = E_s d$, at which point the polarization current peak reverts back to the background signal, is plotted in the main pane of Figure 3. In the higher temperature ferroelectric phase ($F_1$ phase), $E_s$ slowly increases from 0.043V/μm to 0.053V/μm, then sharply increases to about 0.090 V/μm at 112°C, corresponding to the temperature range where DSC shows a broad peak (onset of a second ferroelectric or $F_2$ phase). Between 109°C and 100°C, $E_s$ is almost constant, but then begins to increase below 100°C (signaling the onset of a $F_3$ phase), which corresponds to a large DSC peak with about 28 $J/g$ transition enthalpy shown in Figure S1. The changes in the behavior of $E_s$, at temperatures consistent with peaks in the DSC data, clearly indicate transitions between three separate ferroelectric phases.

We also measured the effective polarization as a function of applied voltage. Results at a few selected temperatures are shown in Figure S4 of ESI. At low voltages the slopes are very small, then they increase at moderate voltages and reach saturation at $V_s$ where the polarization, i.e., the switching is completed. This behavior indicates a weak antiferroelectric type arrangement at low voltages, in agreement with the presence of stripes with alternating polarization direction, as shown in Figure 2(g-i).

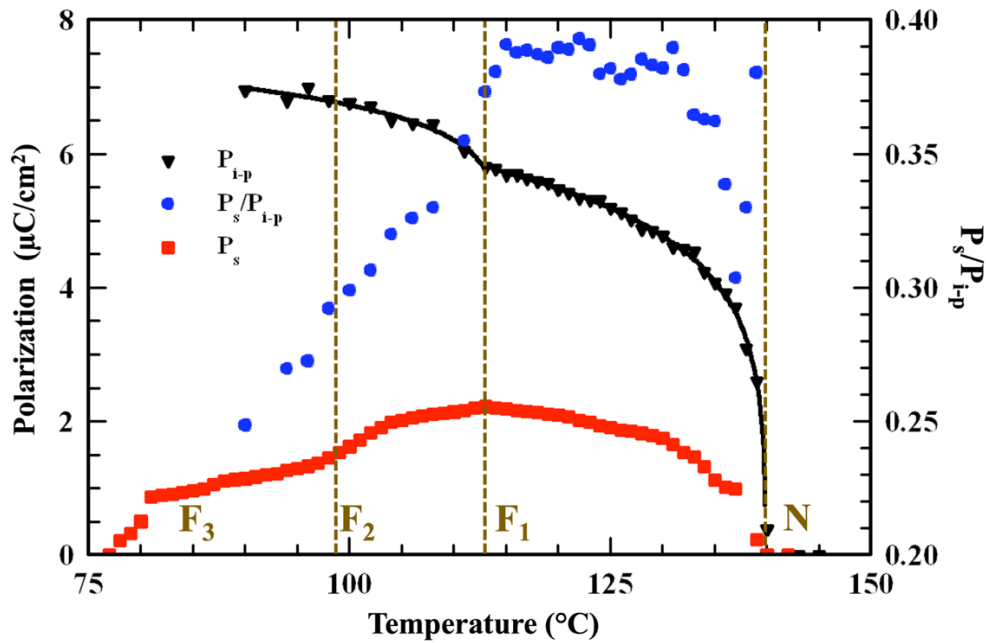

*Figure 4: Summary of polarization measurements using triangular wave fields. Left axis: Temperature dependence of the spontaneous polarization ($P_{i\text{-}p}$) measured using in-plane field (black solid triangles) and in sandwich cell ($P_s$, solid red squares). Right axis: Temperature dependences of the ratio of $P_{i-p}/P_s$.*

The temperature dependence of the ferroelectric polarization (corresponding to the area under the peak in the electric current, see shaded area in the inset of Figure 3) is shown in Figure 4 by black solid triangles. The polarization continuously increases from zero at 140°C to about 5.8μC/cm² at 112°C, then increases further and reaches a value of 6.9 μC/cm² at 90°C. This value is larger than any previously reported on materials purported to be $N_F$.[14,9]

The value of the spontaneous polarization (the dipole density) of RT11001 can be estimated from the molecular dipole moment ($\mu$ = 11.54 Debye), the polar order parameter ($S$), the density ($\rho \sim 1.3 g/cm^3$)[14] and the molar mass ($M$) as $P = S \cdot \mu \cdot \rho \cdot N_A / M$. These values provide $P = S \times 7.2 \frac{\mu C}{cm^2}$; with this we estimate $S \approx 0.95$ at $90°C$ at $E = 0.100 V/\mu m$.

Red solid squares in Figure 4 show the temperature dependence of the saturated polarization $P_s$ measured in the sandwich cell geometry with planar alignment. When the field between the substrates rotates the polarization toward the field by an angle $\theta$, the polarization vector has a component $P_\perp = P_o \cdot sin\theta$ that terminates at the substrates. Consequently, an internal electric field $E_{in} = \frac{P_o sin\theta}{\varepsilon_o(\varepsilon_\perp cos^2\theta + \varepsilon_\parallel sin^2\theta)}$ builds up, where $\varepsilon_\parallel$ and $\varepsilon_\perp$ are the dielectric constants parallel and perpendicular to the director. This field acts against the externally applied field $V/d$ that has to overcome the maximum of the internal field $E_{in}^{max} = \frac{P_o}{2\varepsilon_o \varepsilon_\perp} \cdot \sqrt{\frac{\varepsilon_\perp}{\varepsilon_\parallel - \varepsilon_\perp}}$ so that $V - E_{in}^{max} \cdot d \geq V_s$, where $V_s$ is the voltage needed to switch the polarization. Assuming $\Delta\varepsilon = \varepsilon_\parallel - \varepsilon_\perp > 10^3 \gg \varepsilon_\perp$ (as found for RM737 and DIO with similar dipole moments) and using the measured polarization $P_o = 7 \cdot 10^{-2} \frac{C}{m^2}$ (see Figure 4), we estimate $E_{in}^{max} = 20 \frac{V}{\mu m}$. For $d = 10\mu m$ thick sandwich cells, this gives $V_{in}^{max} = 200V$, which is larger than we could apply without causing formation of air bubbles and burn damage to the samples. Solid blue circles plotted against the right axis show the temperature dependence of the ratio $R = sin\theta(T)$ of the polarizations measured in the sandwich and the in-plane cells. Below the $N$-$F_1$ phase transition $R$ increases sharply up to $R \sim 0.4$. Upon the $F_1$-$F_2$ phase transition $R$ decreases to $R \approx 0.3$ at 100°C and the onset of the $F_3$ phase, where $R$ further decreases to $\approx 0.25$. This temperature dependence of $R$ is consistent with the increasing switching fields in the $F_2$ and $F_3$ phases (Figure 3) thus leading to decreased switching angle $\theta$.

Electric current measurements in sandwich cells can also provide estimates of the electric conductivity; as calculated from the area beneath the polarization current (see Figure S5 of ESI), the conductivity varies between $10^{-6} S/m$ at 139°C and $5 \cdot 10^{-7} S/m$ at 100°C. From the capacitive current at the sign inversion of $dV/dt$ (see Figure S5 of ESI) the effective dielectric constant $\varepsilon_{eff} = \varepsilon_\perp cos^2\theta + \varepsilon_\parallel sin^2\theta$ was found to be about 200. From this (in the approximation that $\Delta\varepsilon \gg \varepsilon_\perp$) we can estimate that $\Delta\varepsilon \sim \frac{200}{0.09} \sim 2 \cdot 10^3$ at 200Hz. This value is comparable to those measured on RM734[13] and DIO.[9]

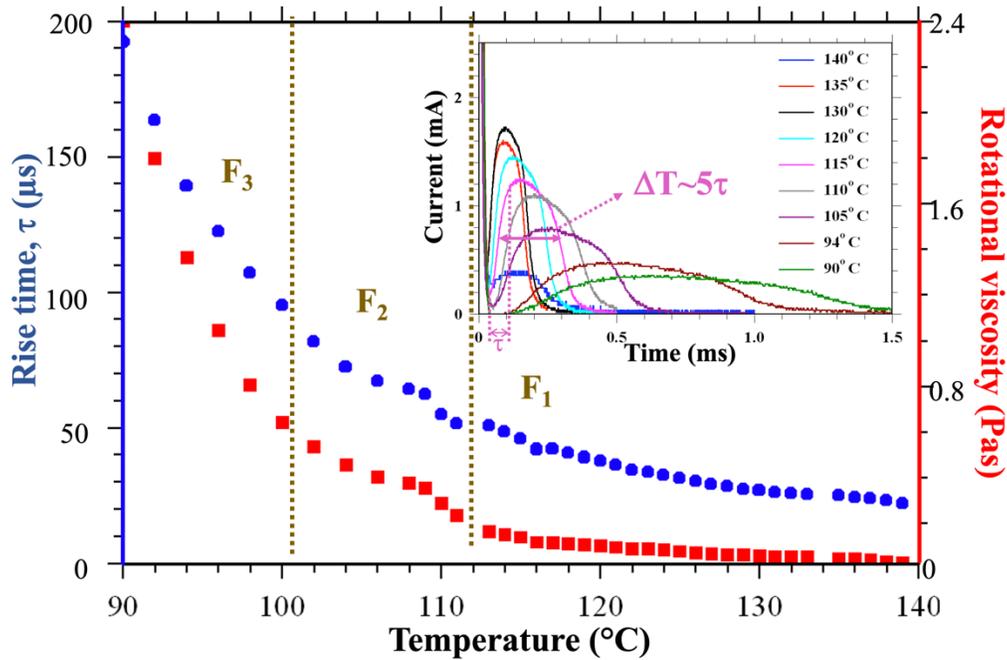

*Figure 5: Temperature dependences of the rise time $\tau = 0.2 \cdot \Delta T$ (blue solid dots plotted against the left axis), and of the rotational viscosity $\gamma_1$ (solid red squares plotted against the right axis). The reversal time ΔT was measured from the full width at half maxima of the time dependence of the electric current flowing through the sample under rectangular electric fields (see inset).*

Lastly, using in-plane cells, we measured the time dependence of the electric current induced by 200 Hz square-wave voltage of $V_{p\text{-}p}$ = 130 V. Examples at various temperatures are shown in the inset to Figure 5. As was found by Chen et al[14], the rise time τ (the time until the polarization current reached 90% of its maximum) is proportional to the polarization reversal time, *ΔT* taken as the FWHM of the polarization current peak (see inset to Figure 5). For RT11001 at 130 $V_{p\text{-}p}$ we find $\tau = 0.2 \cdot \Delta T$. The temperature dependence of τ is plotted against the left axis of Figure 5. On cooling from 140°C to 90°C, the rise time increases from 20μs to 200μs. The curve does show changes in the slopes at the same temperatures where DSC peaks occur at 112°C and 100°C (see Figure 1).

The rotational viscosity $\gamma_1$ is calculated from the rise time as $\tau = \gamma_1/P_oE$ and its temperature dependence is plotted against the right axis of Figure 5. In the $F_1$ phase between 140°C and 112°C the viscosity increases monotonically from 10 mPas to 100 mPa·s, which is typical for low molecular weight nematic fluids. The increase becomes more pronounced in the $F_2$ phase below 112°C, when it increases to about 0.4 Pa·s. In the $F_3$ phase, below 100°C the slope $d\gamma_1/dT$ increases to about *0.2* Pa·s/K , possibly indicating the onset of translational order.

4. **Conclusions**

These experimental results clearly indicate that RT11001 exhibits three distinct ferroelectric phases ($F_1$, $F_2$, $F_3$) below the ordinary nonpolar nematic phase. The fact that polarizing optical microscopy and birefringence measurements do not show observable differences between the three ferroelectric phases strongly suggests that transitions between these phases do not correspond to changes in the director field, but rather to changes in molecular correlations normal to the director. The temperature dependence of the density and spontaneous electric polarization clearly show the $F_1$ to $F_2$ transition, but not the $F_2$ to $F_3$. This corresponds to an increase in the dipole density in the plane perpendicular to the polarization vector, which is manifested as an increase of both the polarization and of the mass density.

The calorimetry, the temperature dependence of the field required to switch the polarization, and the rotational viscosity all show different behavior in all three ferroelectric phases. The rotational viscosity and the switching field show stepwise increases at the $F_1$-$F_2$ transition, in accordance with the conclusion above that the molecular packing normal to the director becomes closer at this transition. One possibility is the formation of short-range spatial correlation between the long axis of the molecules. The strong increase of the rotational viscosity and of the switching field below the $F_2$-$F_3$ transition indicates longer range spatial correlations normal to the director. This is also in accordance with the occurrence of the large DSC peak starting below 100°C corresponding to broad transition between a 3-D anisotropic fluid with short positional correlation (nematic phase) to a 1-D fluid with 2-D positional order (columnar phase). Such combinations of the orientational and positional order in the $F_1$, $F_2$ and $F_3$ phases are consistent with the following proposed scenario: $F_1$ is a purely orientationally-ordered ferroelectric nematic phase ($N_F$), $F_2$ is a ferroelectric nematic phase with short range hexagonal order normal to the director ($N_{hF}$), and $F_3$ has long-range hexagonal order normal to the director ($Col_{hF}$).

An alternative framework is suggested by the spontaneous splay deformation described by Mertelj et al[12], Sebastian et al[13], and Rosseto and Selinger[23]. Accordingly, we could envision a proper non-modulated ferroelectric nematic ($N_F$) structure in the $F_1$ phase, as proposed by Chen et al[14], a one-dimensionally splayed ferroelectric nematic ($N_{SF}$) structure for the $F_2$ phase and a two-dimensionally splayed nematic ($N_{2SF}$) structure for the $F_3$ phase. Such a phase sequence is consistent with the increasing switching threshold and rotational viscosity. However, within this realm one would expect to see a decrease of the birefringence upon an $N_F$ – $N_{SF}$ transition and also upon the $N_{SF}$ – $N_{2SF}$ transition, which is not found in RT11001. For this reason, we suggest that the $N_F$ - $N_{hF}$ - $Col_{hF}$ phase sequence on cooling is a more reasonable conjecture. Future SAXS/WAXS measurements on RT11001 are planned to verify this sequence.

## 5. Acknowledgement

This research was supported by the National Science Foundation under grant DMR-1904167.

# Electronic Supplementary Information

## Multiple ferroelectric nematic phases of a highly polar liquid crystal compound

Rony Saha[1,2], Pawan Nepal[3], Chenrun Feng[2,4], Md Sakhawat Hossein[4], James T. Gleeson[1], Samuel Sprunt[1,2], Robert J. Twieg[3], Antal Jákli[1,2,4]

[1]Department of Physics
[2]Advanced Materials and Liquid Crystal Institute
[3]Department of Chemistry and Biochemistry
[4]Materials Science Graduate Program

Kent State University, Kent, OH 44242, USA

## Contents

## I. Materials and Analytical Hardware

All the solvents used were obtained from commercial source and were used without further purification. Unless specifically stated, all solvents used for the reactions were high purity anhydrous solvents. The starting materials and reagents were obtained from a variety of commercial sources: Acros (oxalyl chloride, Oxone); Alfa-Aesar (DCC); Ambeed (2-fluoro-4-hydroxybenzaldehyde); Combi-blocks (2,4-dimethoxybenzoic acid); Oakwood (DMAP, 4-nitrophenol). Product purifications were done by column chromatography using silica gel (Silicycle 60-120 mesh) and/or by recrystallization from analytical grade solvents.

Polarized optical microscopy (POM) was run on a Nikon ECLIPSE E600 Microscope & SPOT™ idea COMS with temperature programming with a Mettler Toledo FP90 central processor and FP82HT hot stage. IR analysis was accomplished by using a Bruker Vector 33 FTIR spectrometer (Bruker Optics Inc., Billerica, MA, USA) and data obtained was processed and plotted using OPUS software (ver. 6.5, Bruker Optics Inc., Billerica, MA, USA). A Bruker 400 NMR was used for NMR data acquisition (Frequency: 400 MHz for $^1$H-NMR; 100 MHz for $^{13}$C-NMR) and the plots were generated by TOPSPIN 2 software (ver. 2.1, Bruker Optics Inc., Billerica, MA, USA). Product purity and reaction progress was monitored by thin layer chromatography (Dynamic Adsorbents 79011 TLC plates). The developed TLC plates were visualized under UV light.

## II. Synthesis

**2,4-Dimethoxybenzoyl chloride**

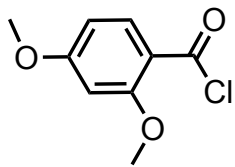

C₉H₉ClO₃.  MW: 200.62   CAS: 39828-35-8

In a 500 ml recovery flask fitted with stirbar, addition funnel and nitrogen bubbler were placed 2,4-dimethoxybenzoic acid (18.1 gm, 100.0 mmol) and dry dichloromethane (75 ml). The resulting slurry was stirred in a cold-water bath and a mixture of oxalyl chloride (19.04 g, 150.0 mmol) in dichloromethane (25 ml) was carefully added over about 90 min (CAUTION: vigorous gas evolution). The bath was allowed to warm to ambient and the mixture was stirred overnight. The next day the stirbar was removed and the solution was concentrated by rotary evaporation. Dry dichloromethane (75 cc) was added as a getter and the mixture was again concentrated by rotary evaporation. On standing at room temperature under vacuum the liquid product solidified (20.22 gm, ~100%). The acid chloride prepared in this fashion is of sufficient purity for further use.

¹HNMR (DMSO-$d_6$, 400 MHz) δ 3.80 (s, 3H), 3.82 (s, 3H), 6.55-6.61 (m, 2H), 7.70 (d, $J$ = 8.6 Hz, 1H).

IR (cm$^{-1}$): 3250, 3127, 3012, 2992, 2952, 2845, 1764, 1722, 1599, 1563, 1504, 1472, 1456, 1434, 1417, 1334, 11297, 1271, 1218, 1200, 1176, 1125, 1019.

### 3-Fluoro-4-formylphenyl 2,4-dimethoxybenzoate

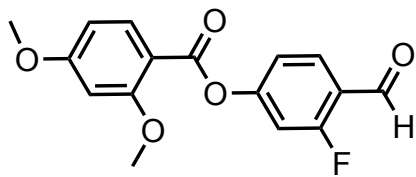

C$_{16}$H$_{13}$FO$_5$   MW: 304.27

In a 500 ml recovery flask was placed 2-fluoro-4-hydroxybenzaldehyde (2.80 gm, 20.0 mmol) and dry dichloromethane (50 ml). The mixture was stirred in a cold-water bath under nitrogen and dry pyridine (6.33 gm, 80.0 mmol) and DMAP (50 mg) were added. When everything dissolved the 2,4-dimethoxybenzoyl chloride (4.00 gm, 20.0 mmol) dissolved in dry dichloromethane (10 ml) was added over about one minute. The mixture was stirred in the water bath for 90 min after which time TLC indicated the presence of a small amount of aldehyde and so additional 2,4-dimethoxybenzoyl chloride (0.20 gm, 5%) was added. After stirring overnight at room temperature, the material was concentrated by rotary evaporation, tetrahydrofuran (50 ml) was added, and the material was concentrated by rotary evaporation again. Next, 5% aqueous HCl (~200 ml) was added dropwise followed by ice and water to nearly fill the flask. The resulting slurry was vigorously stirred to break it up and then the solid was isolated by suction filtration, washed well with water and air dried. The product which was pure enough for subsequent oxidation was obtained as a tan solid (5.32 gm, 88 %).

MP = 129.1 °C.

$^1$HNMR (DMSO-$d_6$, 400 MHz) δ 3.88 (s, 3H), 3.89 (s, 3H), 6.67-6.73 (m, 2H), 7.26-7.31 (m, 1H), 7.43-7.47 (m, 1H), 7.91-7.95 (m, 1H), 7.99 (d, $J$ = 8.7 Hz, 1H), 10.20 (s, 1H).

$^{13}$CNMR (DMSO-$d_6$, 100 MHz) δ 56.2, 56.5, 99.4, 106.3, 109.7, 111.5, 111.8, 119.7 (d), 121.9 (d), 130.8 (d), 134.7, 157.0, 157.1, 162.3, 162.5, 162.8, 165.4, 165.7, 187.4, 187.5.

$^{19}$FNMR (DMSO-$d_6$, 376 MHz) δ -118.39 (m, 1F).

IR (cm$^{-1}$): 3105, 3076, 3045, 3019, 2955, 2887, 2849, 1741, 1696, 1605, 1578, 1605, 1578, 1510, 1500, 1477, 1437, 1410, 1341, 1298, 1274, 1220, 1196, 1153, 1132, 1113, 1012.

### 3-Fluoro-4-carboxyphenyl 2,4-dimethoxybenzoate

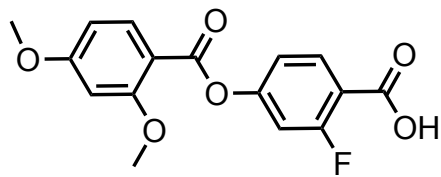

$C_{16}H_{13}FO_6$  MW: 320.27

In a 500 ml round bottom flask with stirbar was placed 3-fluoro-4-formylphenyl 2,4-dimethoxybenzoate (4.56 gm, 15.0 mmol) and DMF (80 ml). The mixture was stirred until everything dissolved and then solid Oxone (9.83 gm, 16.0 mmol) was added all at once. After stirring at room temperature for six hours TLC indicated that the aldehyde was consumed and 5% HCl (200 ml) was added dropwise with stirring followed by ice and water to fill the flask. The resulting slurry was vigorously stirred for an hour and then the solid was isolated by suction filtration, washed well with water and air dried on the filter overnight. The product obtained (4.86 gm, ~100%) still retained some water so it was stored under vacuum in a desiccator over $P_2O_5$ prior to use in subsequent esterification reactions.

MP = 109.3 °C.

$^1$HNMR (DMSO-$d_6$, 400 MHz) δ 3.87 (s, 3H), 3.88 (s, 3H), 6.67-6.73 (m, 2H), 7.19-7.21 (m, 1H), 7.30-7.35 (m, 1H), 7.93-7.99 (m,2H).

$^{13}$CNMR (DMSO-$d_6$, 100 MHz) δ 56.2, 56.5, 99.4, 106.2, 109.9, 111.7, 112.0, 117.0, 117.1, 118.9, 119.0, 133.2, 134.6, 155.3, 155.4, 160.7, 162.4, 162.5, 163.2, 164.9, 165.6.

$^{19}$FNMR (DMSO-$d_6$, 376 MHz) δ -107.63 (m, 1F).

IR (cm$^{-1}$): 2988, 2950, 2845, 1743, 1688, 1607, 1572, 1507, 1418, 1327, 1297, 1239, 1208, 1140, 1125, 1095, 1053, 1027, 1010.

### 4-Nitrophenyl 4-[(2,4-dimethoxylbenzoyl)oxy]-2-fluorobenzoate (RT11001)

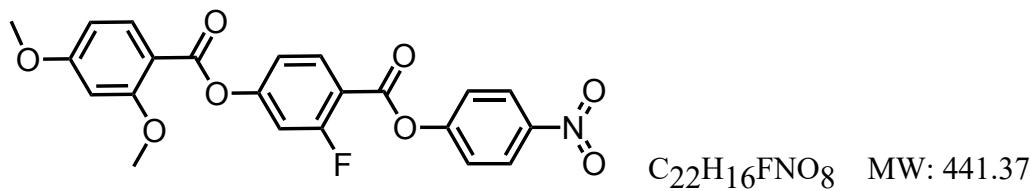

$C_{22}H_{16}FNO_8$  MW: 441.37

In a 500 ml recovery flask fitted with stirbar and nitrogen inlet was placed 3-fluoro-4-carboxyphenyl 2,4-dimethoxybenzoate (0.960 gm, 3.0 mmol), dry dichloromethane (60 ml), 4-nitrophenol (0.414 gm, 3.0 mmol) and DMAP (0.015 gm, 3.75%) The resulting mixture was stirred at room temperature and then DCC (0.824 gm, 4.0 mmol, 1.33 eq) was added all at once. The initial mixture cleared, and a new precipitate soon appeared. After stirring for six hours TLC indicated that no carboxylic acid starting material was evident and so the reaction was terminated. Silica gel (40 cc) and ethyl acetate (40 ml) were added, and the slurry was concentrated to dryness by rotary evaporation. This adsorbed material was placed at the top of a silica gel column (~ 5 cm x 25 cm) made up with dichloromethane and eluted with dichloromethane. Fractions containing only the desired product were combined, concentrated and then recrystallized from 1-propanol to provide the desired product (0.459 gm, 36 %).

Phase sequence in cooling:

$$Iso - 161°C - N - 140.3°C - N_F - 112°C - F_2 - F_3 - 70°C - Cr$$

$^1$HNMR (DMSO-$d_6$, 400 MHz) δ 3.89 (s, 3H), 3.90 (s, 3H), 6.69-6.74 (m, 2H), 7.34-7.36 (m, 1H), 7.49-7.53 (m, 1H), 7.66 (d, $J$ = 9.2 Hz, 2H), 8.01 (d, $J$ = 8.8 Hz, 1H), 8.20-8.25 (m, 1H), 8.38 (d, $J$ = 9.2 Hz, 2H).

$^{13}$CNMR (DMSO-$d_6$, 100 MHz) δ 56.2, 56.5, 99.5, 106.3, 109.6, 112.2, 112.4, 114.6 (d), 119.4, 119.5, 123.9, 125.8, 133.9, 134.7, 145.8, 155.6, 156.7, 156.9, 161.1 (d), 161.2, 162.3, 162.5, 163.8, 165.7.

$^{19}$FNMR (DMSO-$d_6$, 376 MHz) δ -106.15 (m, 1F).

IR (cm$^{-1}$): 3109, 3076, 2981, 2948, 2850, 1745, 1612, 1573, 1519, 1461, 1416, 1353, 1324, 1306, 1262, 1233, 1207, 1152, 1118, 1041, 1010.

## III. Analysis (NMR, IR)

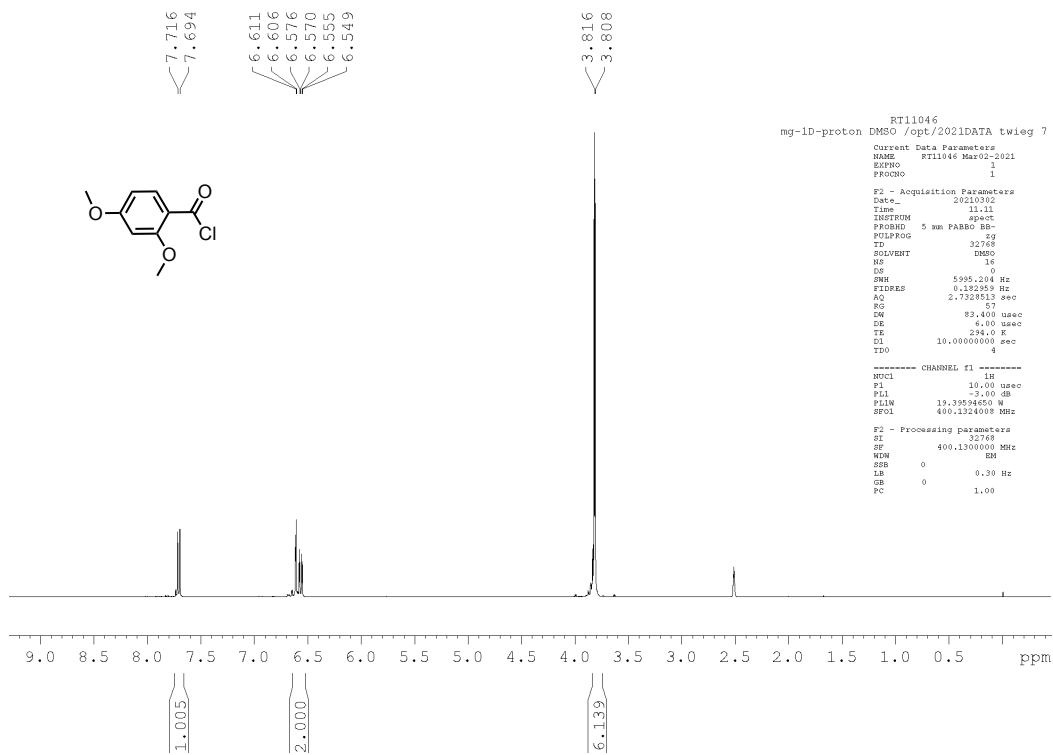

$^1$HNMR of 2,4-Dimethoxybenzoyl chloride

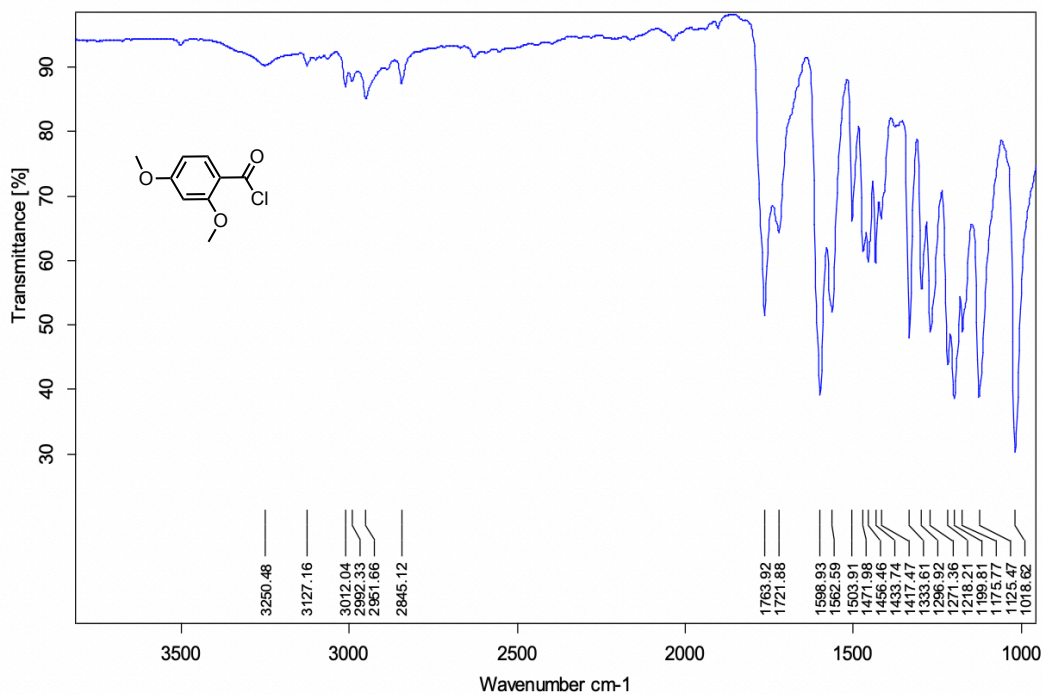

IR of 2,4-Dimethoxybenzoyl chloride

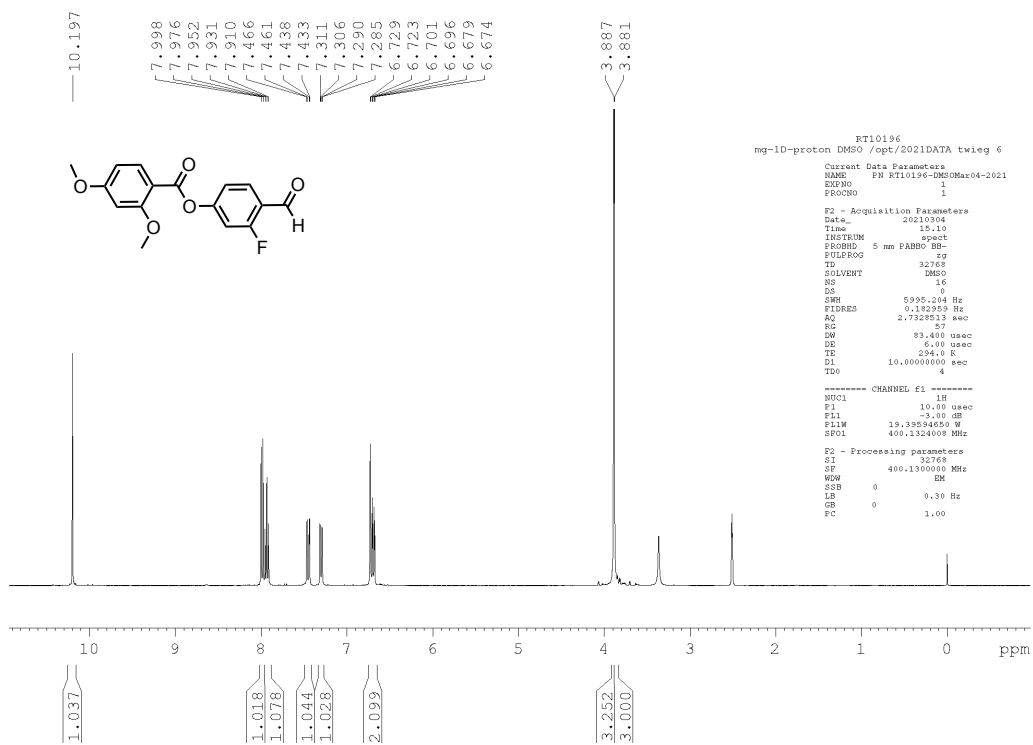

¹HNMR of 3-Fluoro-4-formylphenyl 2,4-dimethoxybenzoate

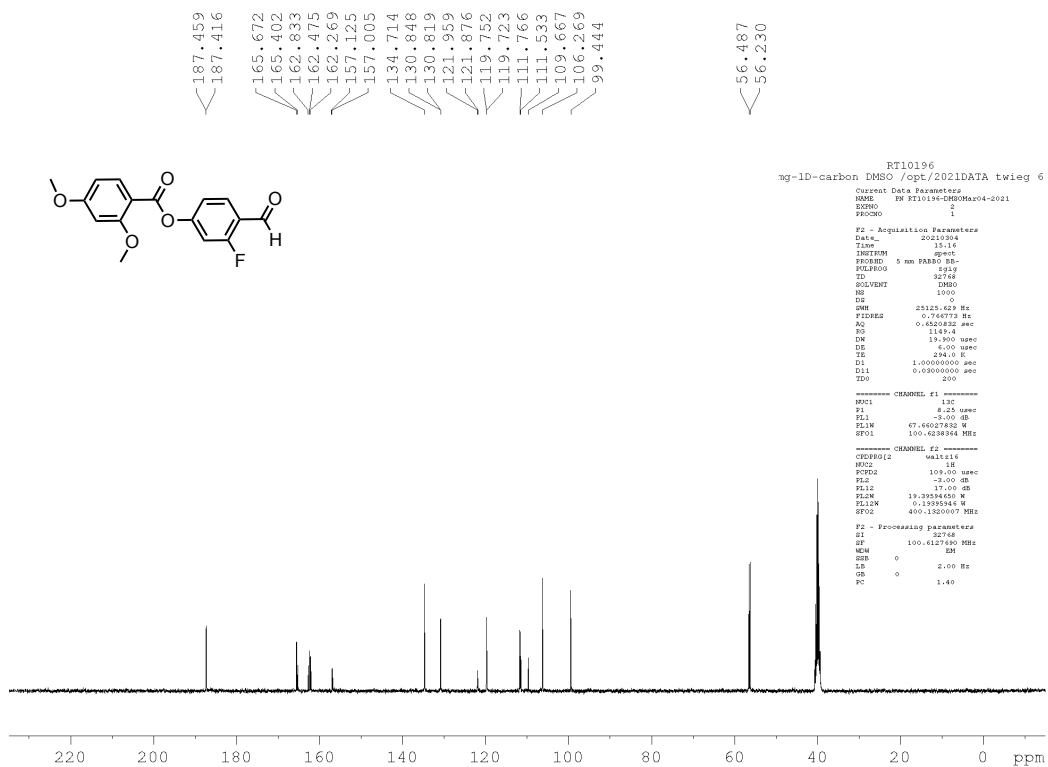

¹³CNMR of 3-Fluoro-4-formylphenyl 2,4-dimethoxybenzoate

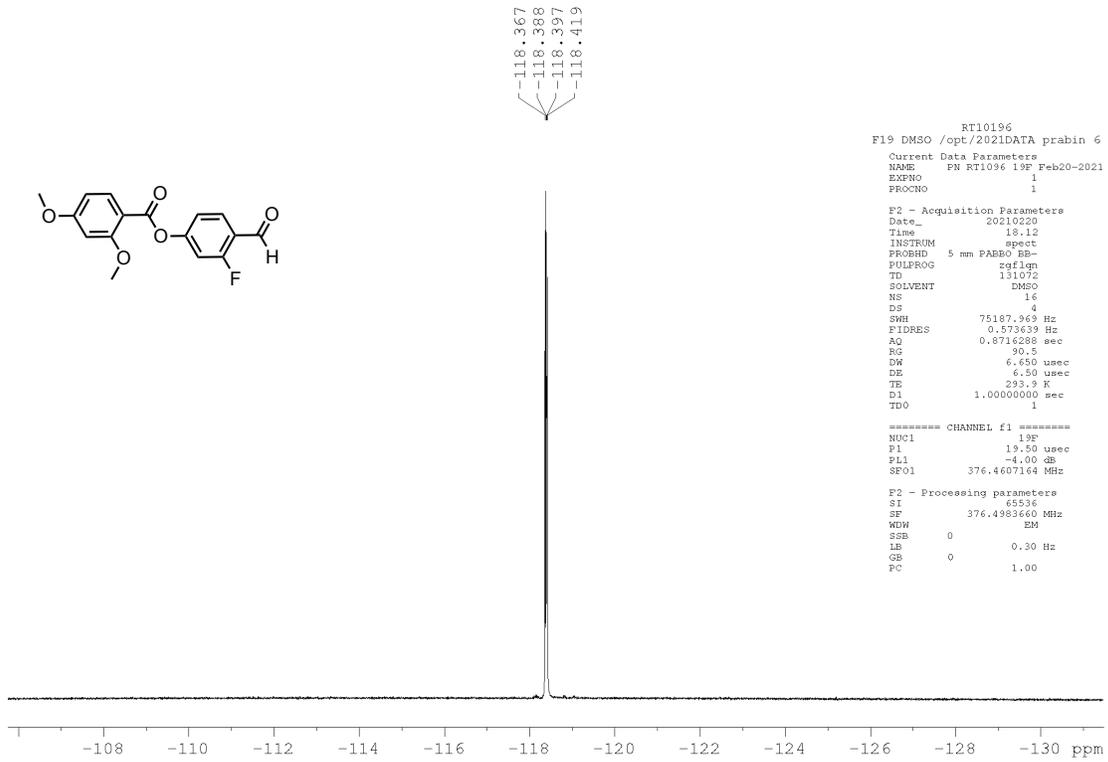

$^{19}$FNMR of 3-Fluoro-4-formylphenyl 2,4-dimethoxybenzoate

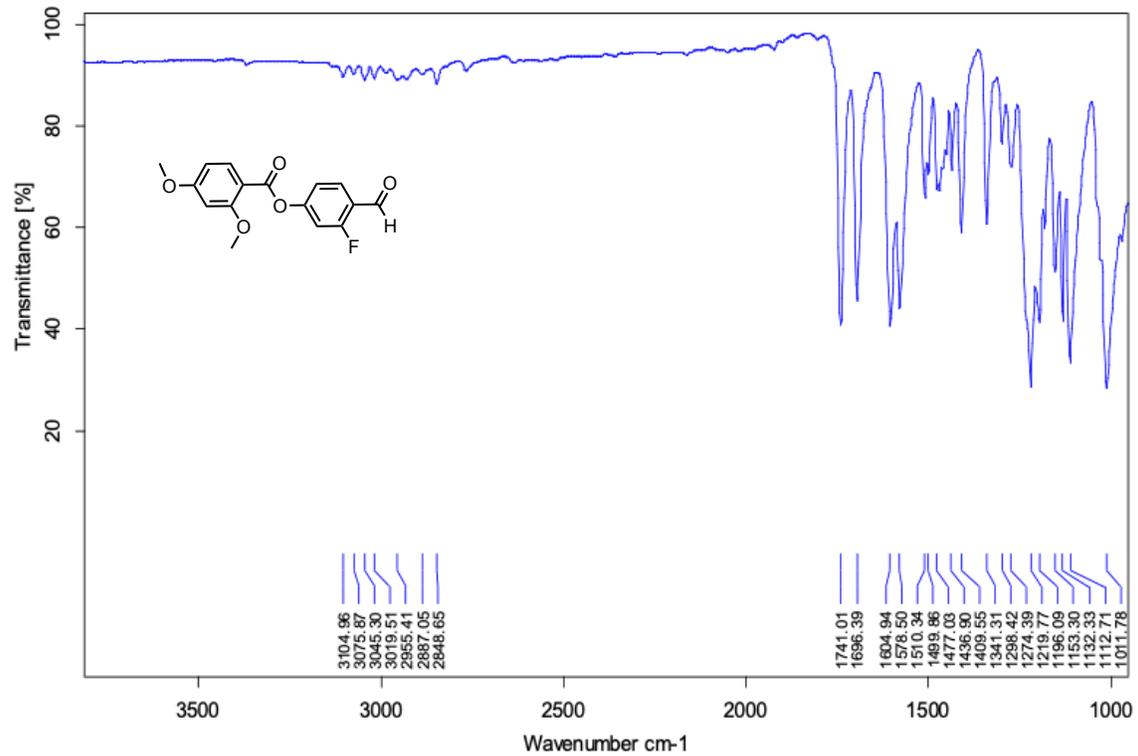

IR of 3-Fluoro-4-formylphenyl 2,4-dimethoxybenzoate

S 9

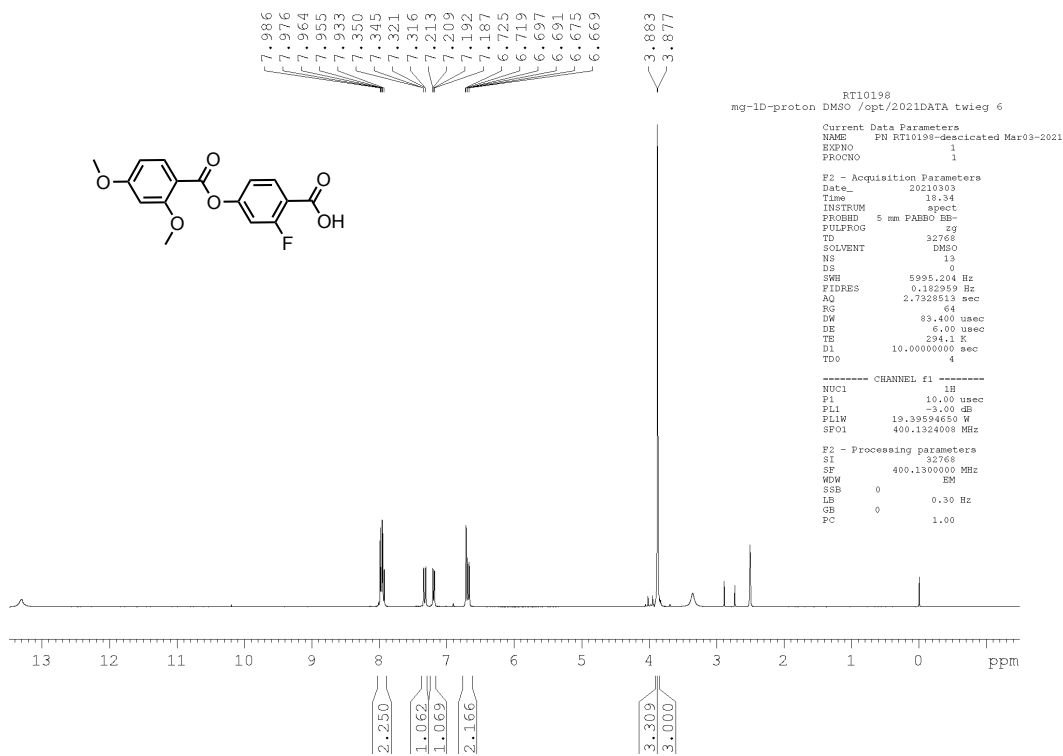

¹HNMR of 3-Fluoro-4-carboxyphenyl 2,4-dimethoxybenzoate

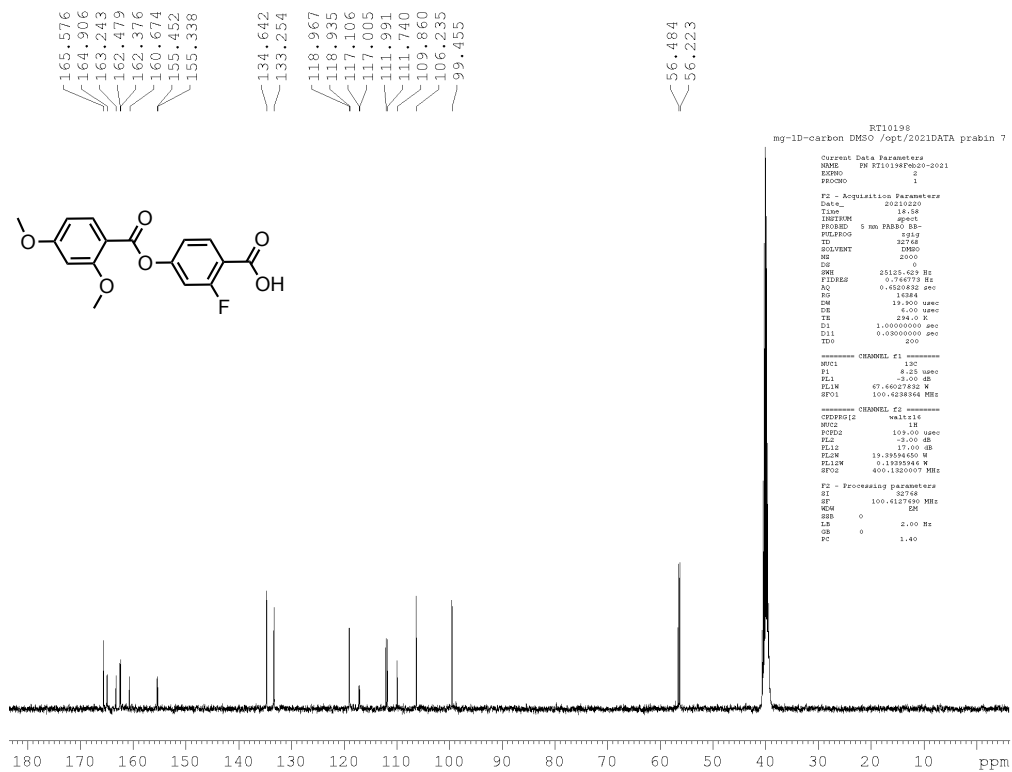

¹³CNMR of 3-Fluoro-4-carboxyphenyl 2,4-dimethoxybenzoate

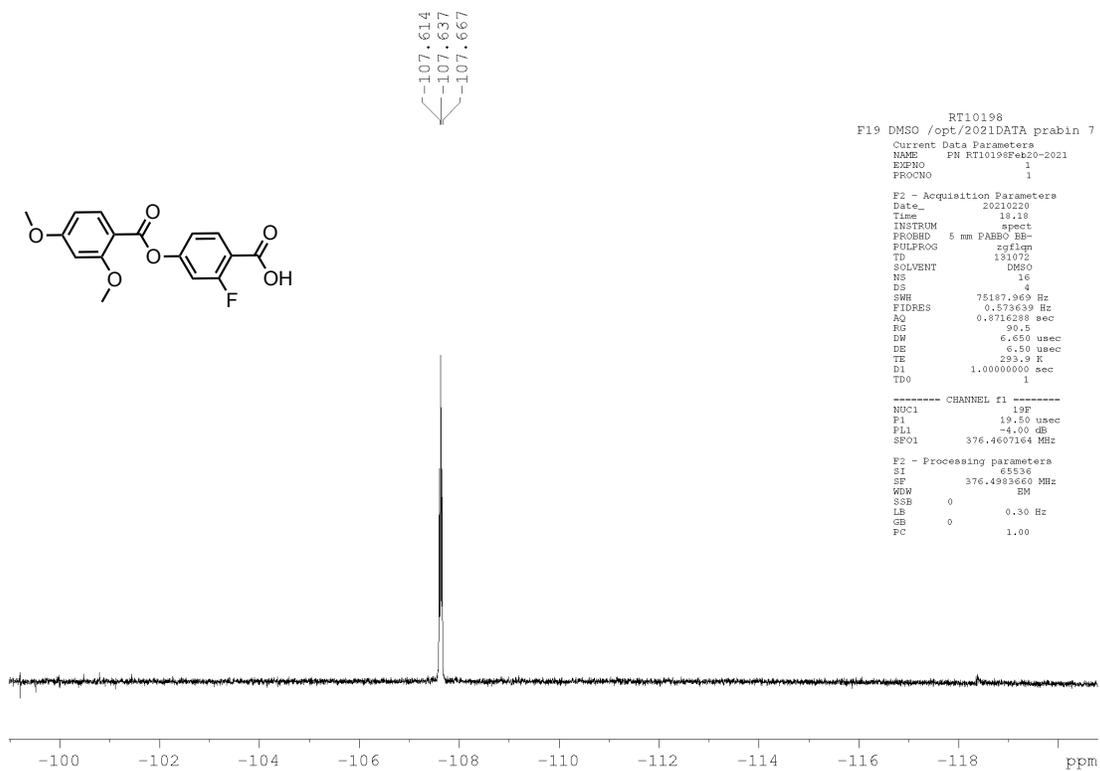

19FNMR of 3-Fluoro-4-carboxyphenyl 2,4-dimethoxybenzoate

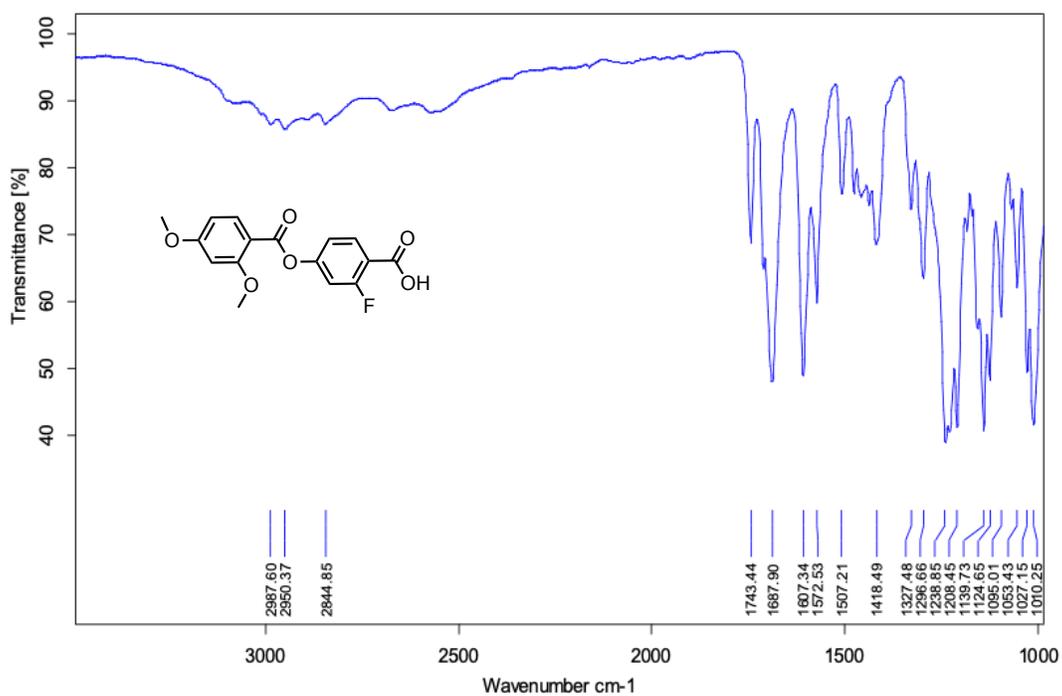

IR of 3-Fluoro-4-carboxyphenyl 2,4-dimethoxybenzoate

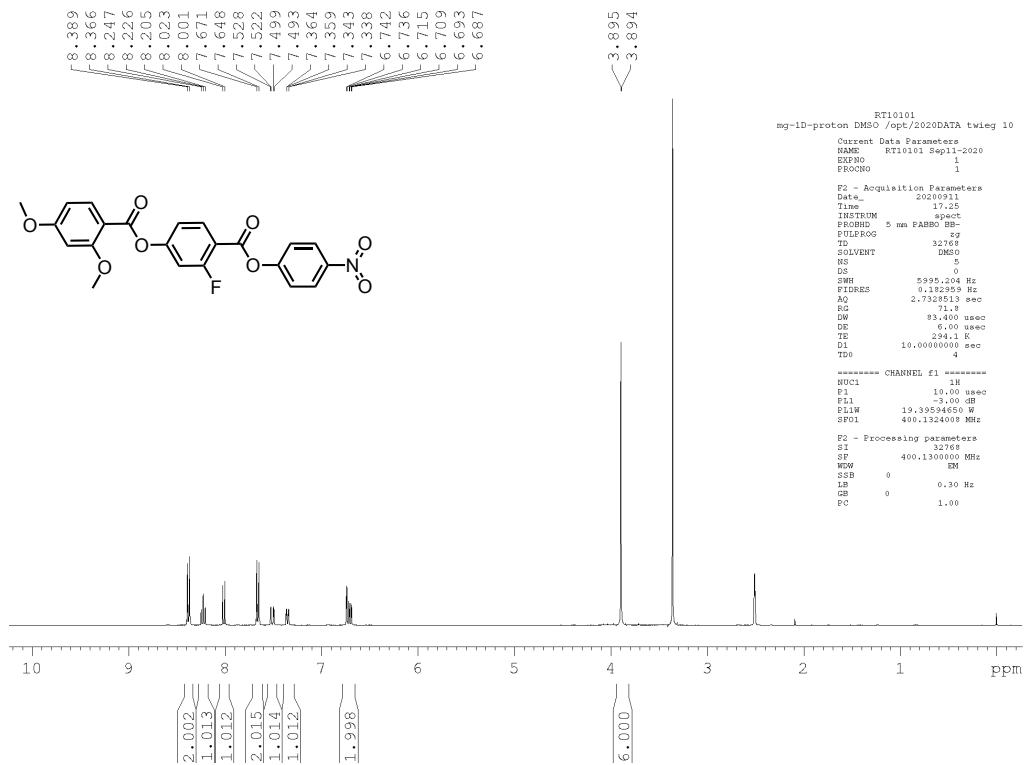

¹HNMR of 4-Nitrophenyl 4-[(2,4-dimethoxylbenzoyl)oxy]-2-fluorobenzoate **(RT11001)**

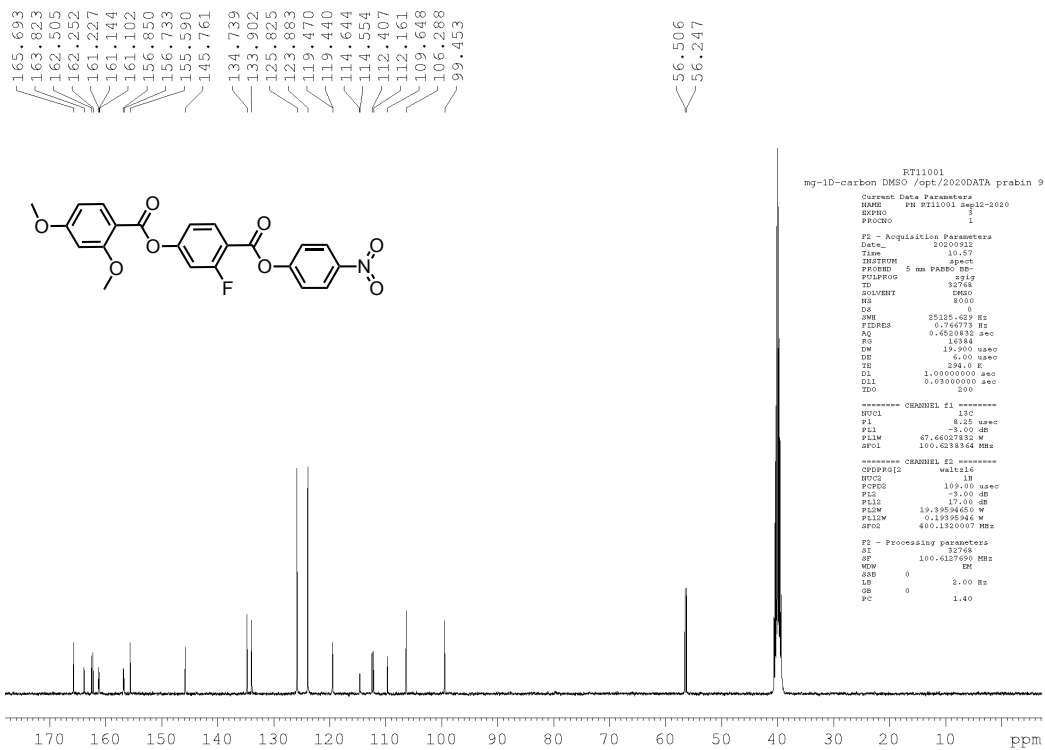

¹³CNMR of 4-Nitrophenyl 4-[(2,4-dimethoxylbenzoyl)oxy]-2-fluorobenzoate **(RT11001)**

S12

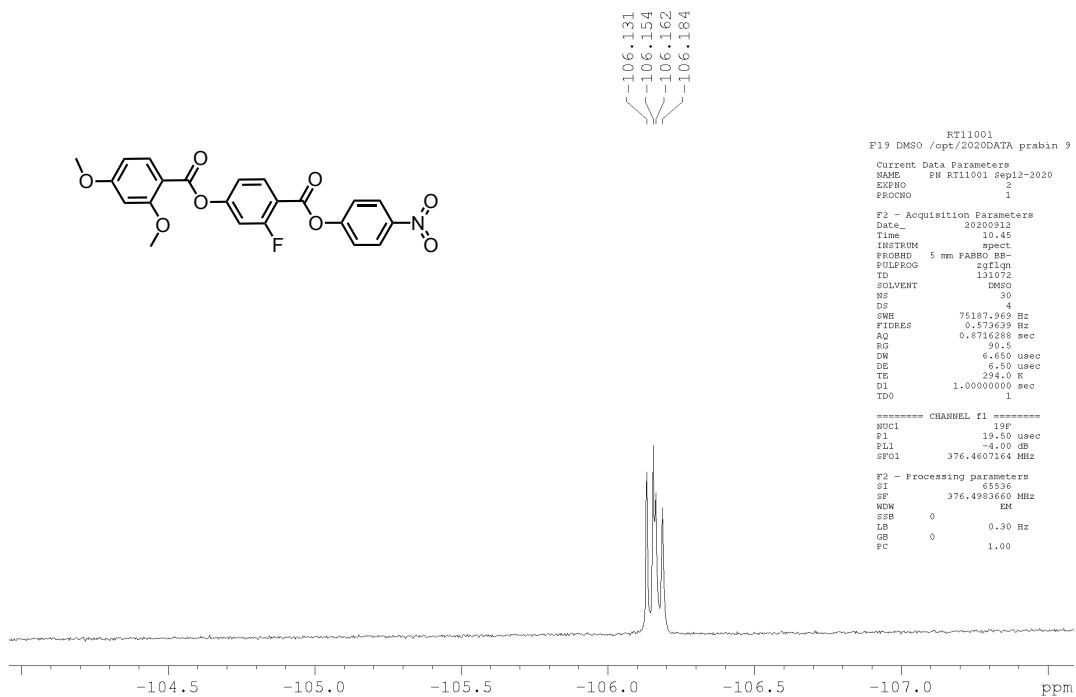

19FNMR of 4-Nitrophenyl 4-[(2,4-dimethoxylbenzoyl)oxy]-2-fluorobenzoate **(RT11001)**

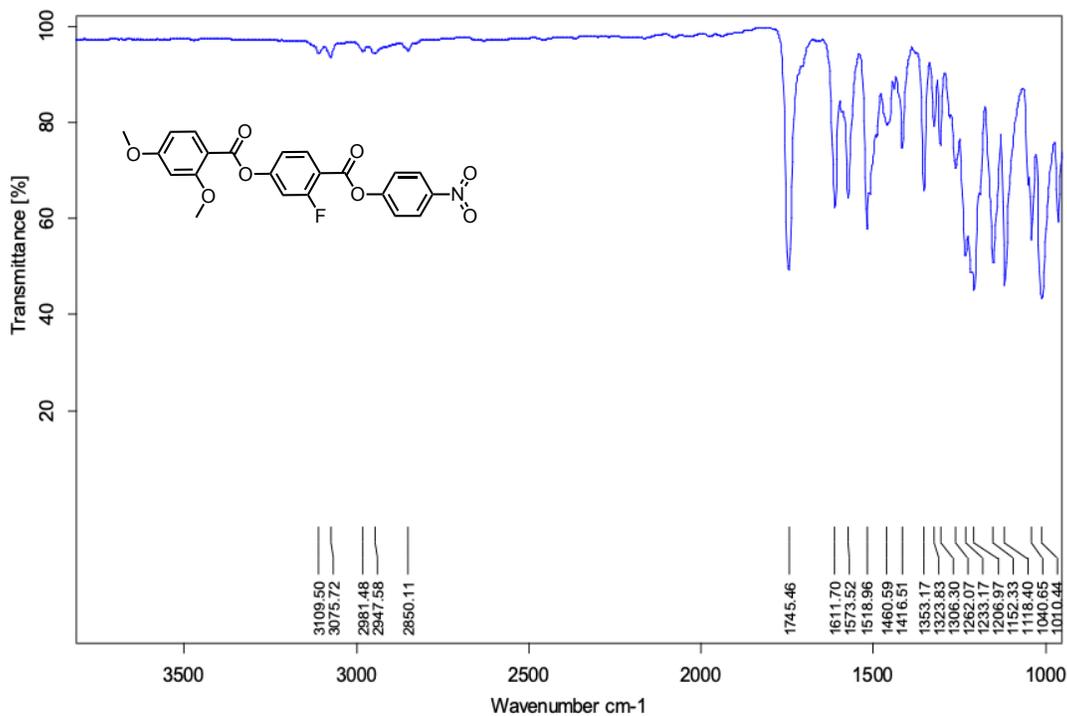

IR of 4-Nitrophenyl 4-[(2,4-dimethoxylbenzoyl)oxy]-2-fluorobenzoate **(RT11001)**

IV. Methods

The DSC scans were run in an aluminum pan under a nitrogen atmosphere.

For polarized Optical Microscopy (POM) measurements, the sample cell was placed in an Instec HS2000 heat stage and viewed through crossed polarizers with an Olympus BX60 microscope.

The birefringence measurements were done using standard photo-elastic modulator techniques at wavelength $\lambda$=632.8 nm. This technique directly measures the phase difference between the ordinary and extra ordinary rays, $\varphi = \frac{2\pi \Delta n d}{\lambda}$, where $\Delta n$ is the birefringence and $d$ is the cell thickness. The data was taken during cooling at 0.2K/min.

Relative density measurements were done by placing the material in a rectangular borosilicate glass capillary with 50µm inner thickness and monitoring the meniscus position as the entire capillary was cooled at 1°C/min. The temperature dependence of the LC-air interface was measured at three points as shown by blue arrows in the inset to Figure 1 and the average values were used to determine the temperature dependence of the density normalized to the density measured in the isotropic phase at 165°C.

Electric current measurements were done using triangular and rectangular waveform electric signals that were generated using a HP 33120A function generator FLC F20AD amplifier. The current was monitored using a $20k\Omega$ resistor in series.

For calculating the molecular dipole moment of RT11001, we used the dispersion-corrected range-separated hybrid $\omega$B97XD functional and the basic set of the software cc-pVDZ[20].

## V. Thermal Properties

1. *DSC traces*

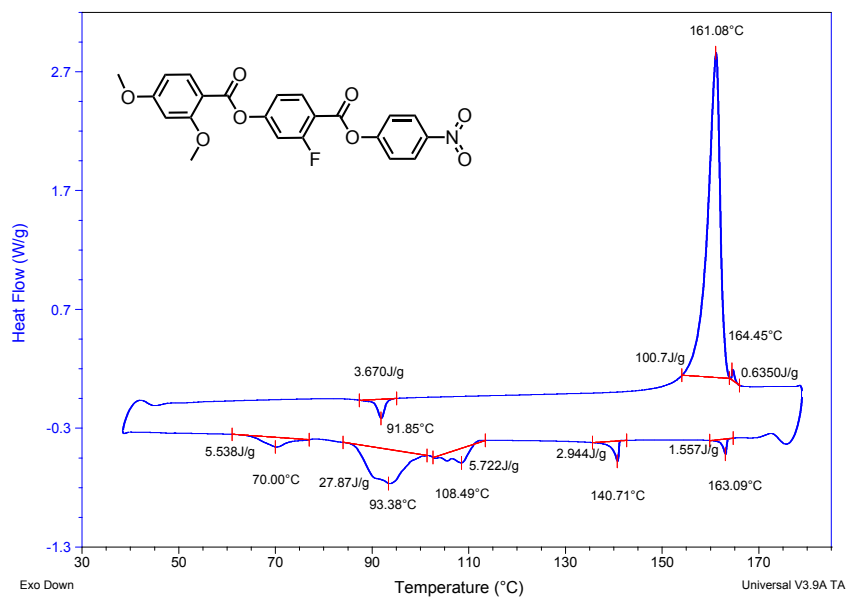

*Figure S1: DSC of **RT11001** before vacuum treatment.*

*2nd heating and cooling cycle, rate = 5 °C/min, sample size = 3.551 mg*

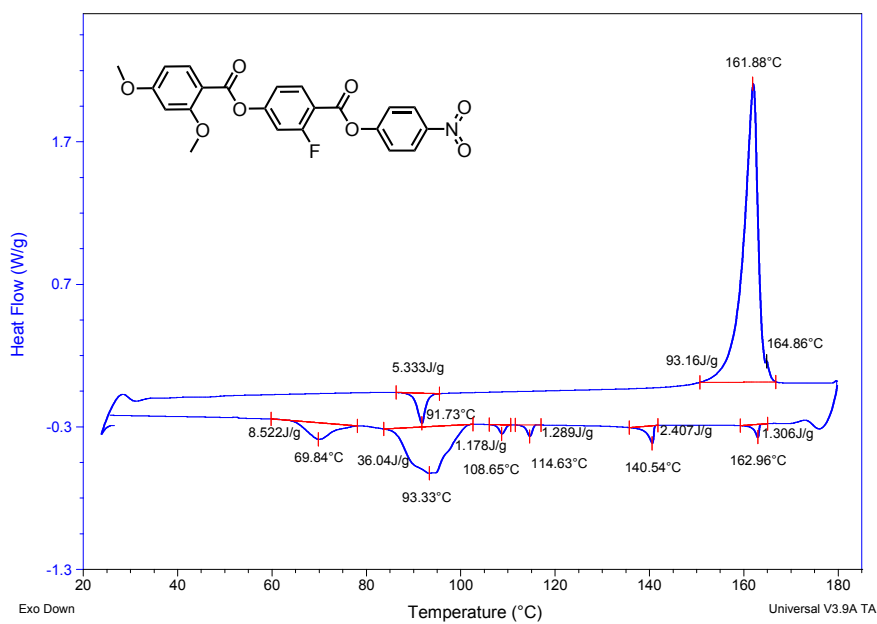

*Figure S2: DSC of **RT11001** after vacuum treatment (100 Torr, ambient T, three days)*

*2nd heating and cooling cycle, rate = 5 °C/min, sample size = 4.890 mg.*

2. *Polarized Optical Textures*

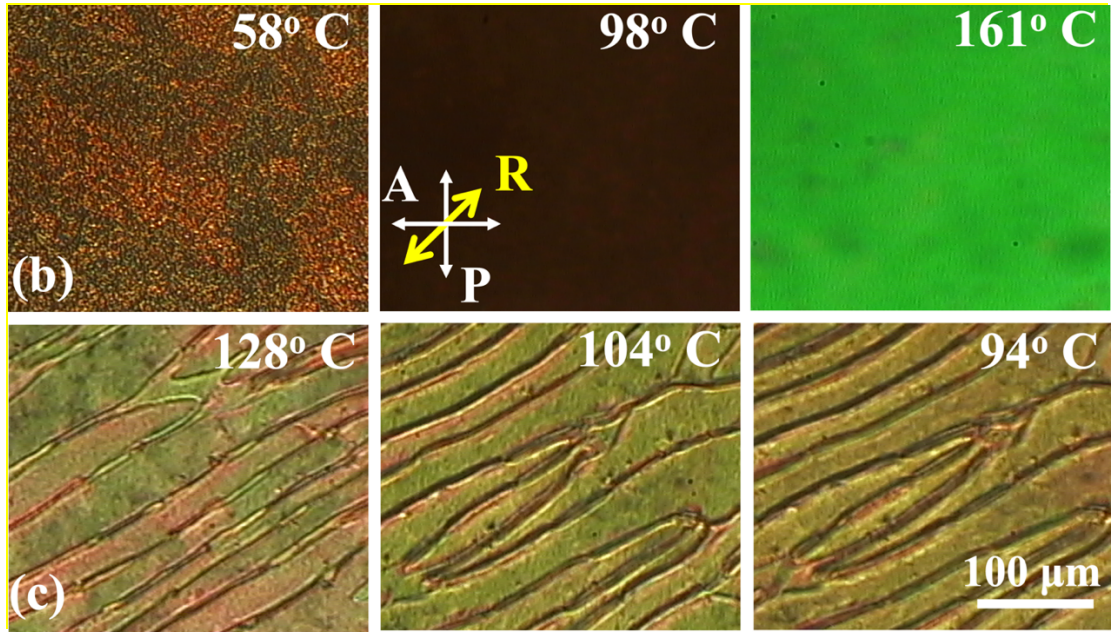

*Figure S3: Representative polarized optical microscopy images of a 10-μm thick cell in planar alignment in heating at 3°C/min rate and in cooling at 1°C /min rate, respectively.*

## VI. Electric current measurements

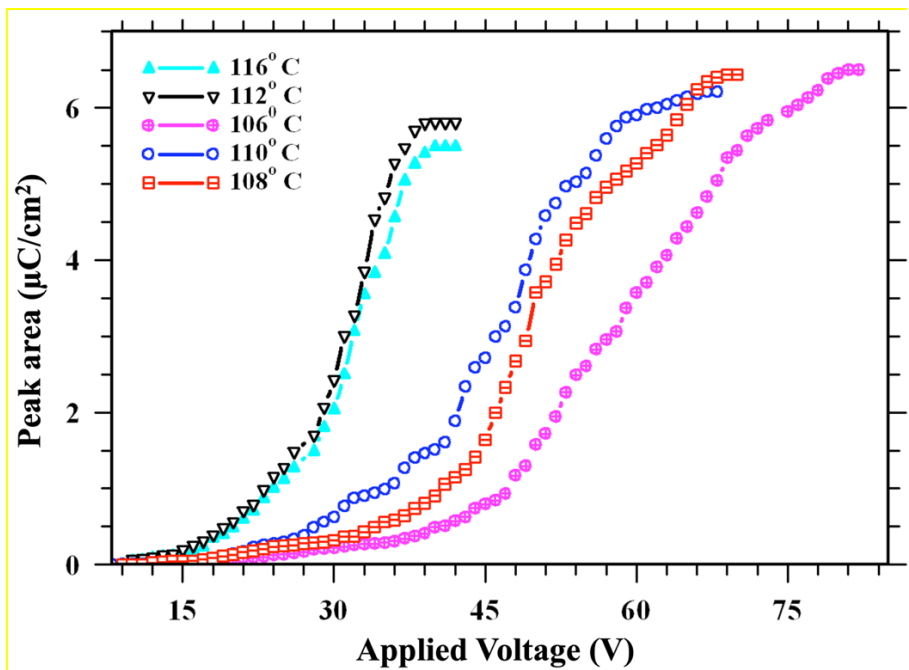

*Figure S4: Voltage dependence of the polarization P calculated from the area beneath the polarization current peak (see peak shown in the inset of Figure 3) at selected temperatures.*

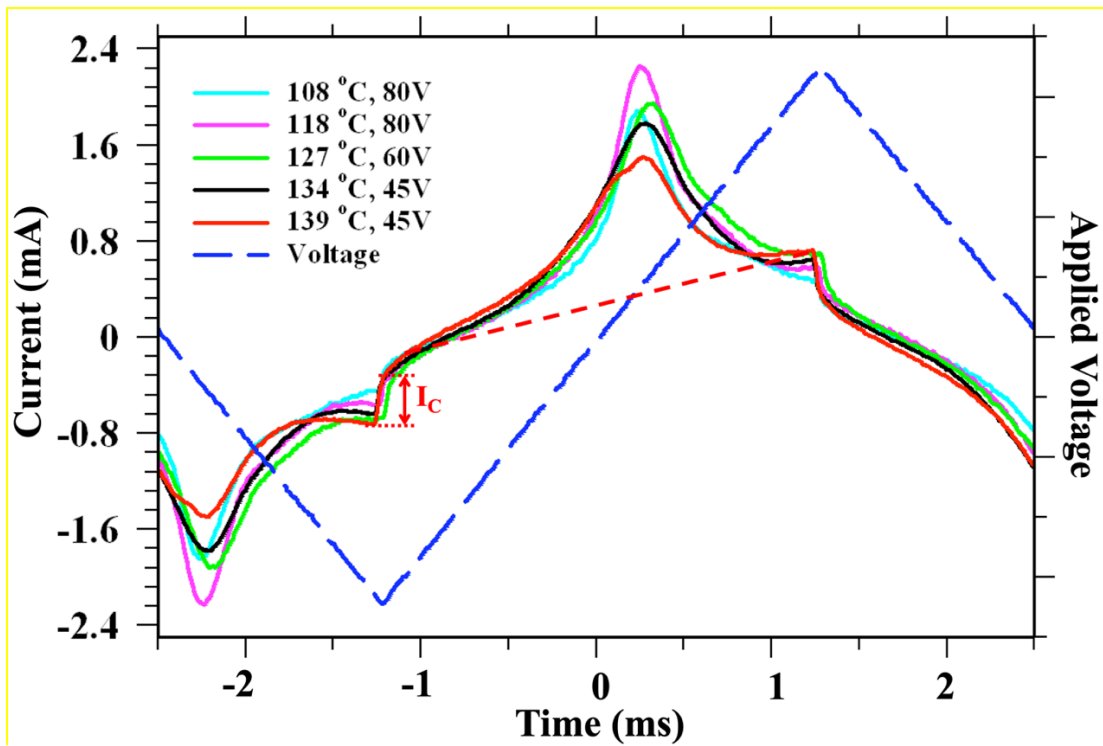

*Figure S5: Time dependences of the electric current flowing through a d=15μm thick film with 1cm² electrode area in sandwich cell geometry. Dotted line illustrates the voltage dependent current through the resistance of the liquid crystal. $I_C$ illustrates the capacitive current upon sign reversal of dV/dt.*